\documentclass[12pt]{JHEP3}

\usepackage{cite}
\usepackage{enumerate}
\usepackage{amsbsy}
\usepackage[dvips]{graphicx}
\usepackage{graphics}

\newcommand{\be}{\begin{equation}} \newcommand{\ee}{\end{equation}}
\newcommand{\bea}{\begin{eqnarray}} \newcommand{\eea}{\end{eqnarray}}

\newcommand{\re}[1]{(\ref{#1})}

\newcommand{\pat}{\partial}

\renewcommand{\sec}[1]{section \ref{#1}}
\newcommand{\fig}[1]{figure \ref{#1}}
\newcommand{\brt}[1]{[#1]}

\renewcommand{\a}{\alpha}
\renewcommand{\b}{\beta}

\newcommand{\LCDM}{$\Lambda$CDM\ }
\newcommand{\GN}{G_{\mathrm{N}}}
\newcommand{\ha}{\frac{1}{2}}

\newcommand{\keq}{k_{\mathrm{eq}}}

\newcommand{\teq}{t_{\mathrm{eq}}}

\newcommand{\rmd}{\mathrm{d}}

\newcommand{\bz}{\bar{z}}

\newcommand{\nonum}{\\}
\newcommand{\etal} {et al.}

\newcommand{\adot}{\dot{a}}
\newcommand{\addot}{\ddot{a}}
\newcommand{\rhodot}{\dot{\rho}}

\newcommand{\HH}{\frac{\adot^2}{a^2}}

\newcommand{\av}[1]{\langle{#1}\rangle}
\newcommand{\sQ}{\mathcal{Q}}
\newcommand{\sR}{{^{(3)}R}}

\newcommand{\Om}{\Omega_{\mathrm{m}}}

\newcommand{\om}{\omega_{\mathrm{m}}}

\newcommand{\PRD}[1]{{\it Phys. Rev.} {\bf D#1}}

\newcommand{\PRL}[1]{{\it Phys. Rev. Lett.} {\bf #1}}

\newcommand{\PLA}[1]{{\it Phys. Lett.} {\bf A#1}}

\newcommand{\MNRAS}[1]{{\it Mon. Not. Roy. Astron. Soc.} {\bf #1}}
\newcommand{\APJ}[1]{{\it Astrophys. J.} {\bf #1}}

\newcommand{\JHEP}[1]{{\it JHEP} {\bf #1}}
\newcommand{\CQG}[1]{{\it Class. Quant. Grav.} {\bf #1}}
\newcommand{\GRG}[1]{{\it Gen. Rel. Grav.} {\bf #1}}
\newcommand{\AaA}[1]{{\it Astron. \& Astrophys.} {\bf #1}}
\newcommand{\PROG}[1]{{\it Prog. Theor. Phys.} {\bf #1}}

\newcommand{\IJMPA}[1]{{\it Int. J. Mod. Phys.} {\bf A#1}}
\newcommand{\IJMPD}[1]{{\it Int. J. Mod. Phys.} {\bf D#1}}

\title{Backreaction: directions of progress}

\author{Syksy R\"{a}s\"{a}nen \\

University of Helsinki, Department of Physics \\
P.O. Box 64, FIN-00014 University of Helsinki, Finland \\

\email{syksy {\it dot} rasanen {\it at} iki {\it dot} fi}}

\abstract{
Homogeneous and isotropic cosmological models with ordinary matter
and gravity predict slower expansion and shorter distances than observed.
It is possible that this failure is due the known breakdown
of homogeneity and isotropy related to structure formation, rather
than new fundamental physics.
We review this backreaction conjecture, concentrating on topics
on which there has been progress as well as open issues.
}

\preprint{HIP-2011-02/TH}

\begin{document}

\setcounter{tocdepth}{2}

\setcounter{secnumdepth}{3}

\section{Introduction} \label{sec:intro}

\subsection{Three choices for a factor of two}

The predictions of homogeneous and isotropic models (with linear
perturbations) with ordinary matter and gravity are off by a factor of
about two compared to observations in the late universe.
Ordinary matter here means that it has non-negative pressure and ordinary
gravity refers to the four-dimensional Einstein-Hilbert action.
The simplest such model, which contains particles of the Standard
Model of particle physics and cold dark matter in a spatially flat
universe,  underpredicts the distance to the last scattering surface
at redshift 1090 by a factor of 1.4--1.7, for a fixed Hubble
parameter today (assuming a power-law spectrum of primordial
perturbations) \cite{Vonlanthen:2010}.
The expansion rate today is wrong by a factor of 2
if we keep the matter density fixed,
$\Omega_\mathrm{m0}\equiv8\pi\GN\rho_\mathrm{m0}/(3H_0^2)\approx0.25$
\cite{Peebles:2004}, or a factor of 1.2--1.5 if we keep the age
of the universe fixed instead, $H_0t_0\approx$ 0.8--1
as opposed to $H_0t_0=2/3$ \cite{Hubble, Krauss:2003}.
This factor of two disagreement means that at least one of
the three assumptions is wrong.
Either there is exotic matter with negative pressure, general
relativity does not hold on cosmological scales, or the
homogeneous and isotropic approximation is not valid at late times.

Mathematically, the simplest possibility is to retain the homogeneous
and isotropic approximation and introduce vacuum energy or
a cosmological constant, which are examples of exotic matter
and modified gravity, respectively.
Quantum field theory predicts that there is an energy density
associated with the vacuum state, so this possibility is theoretically
on very solid ground (unlike most other exotic matter
or modified gravity proposals).
This $\Lambda$CDM model agrees well with most observations, though there
appear to be discrepancies in the distribution of matter on large scales
\cite{inhom, Einasto, Hoyle:2010, Murphy:2010}.
Its main problem is generally considered to be the fact that
the value of the vacuum energy required to explain the observations
is very small compared to known particle physics scales,
$\rho_\mathrm{vac}\approx (2.3$ meV)$^4$.
However, there is no prediction for the vacuum energy, only
arguments based on naturalness. It can even be argued that
the meV scale is quite natural from the point of view of electroweak
physics, as follows.
In the axiomatic approach to quantum field theory in curved
space \cite{axiom}, the vacuum energy of a free scalar
field vanishes in the limit of zero mass.
Let us assume that the same is true for gauge fields and fermions,
and that the Higgs is a composite, so that there are no
fundamental scalar fields.
At high energies, when the electroweak symmetry is unbroken, the
vacuum energy would be zero, while a non-zero value would be generated
in the electroweak phase transition. Naively, we would expect this
energy to be of the order of the electroweak scale. However, it
is expected that in an interacting theory vacuum expectation values
depend non-analytically on the coupling constants \cite{axiom}.
We can make the simple estimate
$\rho_\mathrm{vac}=v^4 e^{-\frac{1}{g^2}}=v^4 e^{-\frac{1}{\alpha}}\approx (1.3 \times 10^{-15} v)^4\approx (0.3$ meV$)^4$,
where $v=246$ GeV is the Higgs vacuum expectation value and $g^2$ is
the coupling, for which we have simply put $\alpha=1/137$.
Substituting for the scale $v$ the sum of particle masses and
taking into account numerical prefactors would change the vacuum
energy by factors of order unity, and it is of course exponentially
sensitive to factors of $4\pi$ and weak mixing angles in the exponent
(and sensitive to the scale at which the coupling is evaluated).
Like arguments in favor of unified scale vacuum energy,
this is little more than inspired numerology, but it shows that
it is not implausible to get the right scale out from the quantum field
theory of electroweak physics.

The vacuum energy density required to explain the observations is
considered problematic not only because of its smallness compared
to known fundamental scales, but also because it has to be close to
the matter density today. Another formulation of this
coincidence problem is that the vacuum energy would have
become important only at late times, when the universe
is about ten billion years old. It seems serendipitous that
we should be witnessing a special and brief dynamical phase
in the evolution of the universe, when the universe is undergoing a
transition from matter domination to vacuum energy domination.
However, the coincidence problem contradicts neither observation
nor any known theoretical law, so at most it provides a
motivation to search for alternatives.

In contrast, a concrete problem with all homogeneous and
isotropic models is that the real universe is far from
homogeneity and isotropy at late times.
Indeed, homogeneous and isotropic models with ordinary
matter and gravity agree well with observations of early times,
and the factor of two disagreement arises in the late universe
when deviations from homogeneity and isotropy become significant.

Before concluding that the introduction of vacuum energy or more
complicated new physics is needed, it is necessary to check
the validity of the homogeneous and isotropic approximation.
The effect of deviations from homogeneity and isotropy
on average quantities (in particular, the average expansion rate)
is called backreaction
\cite{Ellis:2005, Rasanen:2006b, Buchert:2007, Rasanen:2010b}.
Physically, the simplest possibility is that the factor
two discrepancy would be  explained by the known breakdown of
the homogeneous and isotropic approximation related to to structure
formation, without any new fundamental physics
\cite{Buchert:2000, Tatekawa:2001, Wetterich:2001, Schwarz:2002, Rasanen, Kolb:2004}.
This may be called {\it the backreaction conjecture}.

The formation of non-linear structures does in fact lead to
deviations in the local expansion rate which are of the order
of the observed discrepancy, and the process has a preferred
time of about ten billion years. The key issue
is how local deviations add up and cancel in the observed
signal, which involves integrals over large scales.
Note that the success of the homogeneous and isotropic model
with vacuum energy shows that the observations can be explained
simply by increasing the expansion rate (and correspondingly
making distances longer), since that is the only cosmological
effect of vacuum energy.
Furthermore, there is no evidence from local physics for
exotic matter or modified gravity, all of the indications
involve integrals over large scales.
The situation is rather different from that of dark matter,
for which there are several independent lines of evidence
\cite{Roos:2010}, including from local physics.
Also, since most observations probe distances, which
involve the expansion rate only via an integral
(for exceptions, see \cite{ages, BAOradial}),
the expansion history can in fact have significant deviations
from the homogeneous and isotropic vacuum model while still
fitting the data.

The problem is well-defined. Given the known particle content
plus a model of dark matter and starting from a nearly
homogeneous and isotropic state at early times, with a
given Gaussian spectrum of perturbations, how do the
matter distribution and geometry evolve in general relativity?
In particular, we want to find how null geodesics are affected,
since most cosmological observations consist of measurements
of photons. The difficulty arises from the complexity of
non-linear evolution in general relativity.
The problem of finding a tractable approximation scheme
is complicated by the fact that it is not immediately obvious
what are the relevant physical effects which have to be included.
At the moment it is not yet clear whether backreaction is
quantitatively important or not, but there has been
important progress in understanding the phenomenon.

In section 2 we discuss the basics of how structures
affect the average expansion rate. In section 3 we illustrate
the issue with a simple toy model which has correct qualitative
features, and then present a semi-realistic model where we
can estimate magnitude of the effect. In section 4 we briefly
discuss light propagation and its relation to the average
expansion rate.
In section 5 we discuss the role of the Newtonian limit of
general relativity and linear perturbation theory.
We conclude in section 6 with a summary.

\section{The expansion rate}

\subsection{Statistical and exact symmetry}

In cosmology, the evolution of the universe is
usually described with the homogeneous and isotropic
Friedmann-Robertson-Walker (FRW) models, with the justification
that the universe appears to be homogeneous and isotropic on large scales.
However, it is important to distinguish between {\it exact}
and {\it statistical} homogeneity and isotropy.
Exact homogeneity and isotropy means that the space has a
local symmetry: all points and all directions are equivalent.
The FRW models are exactly homogeneous and isotropic.
Statistical homogeneity and isotropy simply means that if we
consider a box anywhere in the universe, the mean quantities
in the box do not depend on its location, orientation or size,
provided that it is larger than the homogeneity scale.
(See \cite{inhom} for a more detailed discussion of statistical
homogeneity and isotropy, and also the issue of self-averaging.)

The early universe is nearly exactly homogeneous and isotropic,
in two ways. First, the amplitude of the perturbations around
homogeneity and isotropy is small.
Second, the distribution of the perturbations is statistically
homogeneous and isotropic. At late times, when density perturbations
become non-linear, the universe is no longer locally near homogeneity
and isotropy, and there are deviations of order unity in quantities
such as the local expansion rate. However, the distribution of the
non-linear regions remains statistically homogeneous and isotropic
on large scales. It has been argued that the homogeneity scale would
have been detected \cite{Hogg:2005}, but the result is disputed
\cite{inhom}; in any case the homogeneity scale is not less than 100 Mpc.
We assume that the universe is indeed statistically homogeneous
and isotropic, with a homogeneity scale much smaller than the Hubble
scale. We are interested in the effects of the structures that are
known to exist, not speculative structures such as Gpc-scale voids.

Due to the statistical symmetry, the average expansion rate evaluated
inside each box is equal (up to statistical fluctuations), but this
does not mean that it would be the same as in a completely smooth
spacetime, because there are structures in the boxes.
We can say that time evolution and averaging do not commute:
if we smooth a clumpy distribution and calculate the time
evolution of the smooth quantities with the Einstein equation,
the result is not the same as if we evolved the full clumpy
distribution and took the average at the end.
Put simply, FRW models describe universes which are exactly
homogeneous and isotropic, not universes which are only
statistically homogeneous and isotropic.
The effect of clumpiness on the average was first discussed
in detail by George Ellis in 1983 under the name
{\it fitting problem} \cite{fitting}.
Clumpiness affects the expansion of the universe, the way light
propagates in the universe and the relationship between the two.
Let us first discuss the expansion rate.

\subsection{The local expansion rate}

We consider a universe where the energy density of
matter dominates over pressure, anisotropic stress
and energy flux everywhere. In other words, the matter
can be considered a pressureless ideal fluid, or dust.
We assume that the relation between the matter and
the geometry is given by the Einstein equation:
\bea \label{Einstein}
  G_{\a\b} &=& 8\pi\GN T_{\a\b} = 8\pi\GN \rho u_{\a} u_{\b} \ ,
\eea

\noindent where $G_{\alpha\beta}$ is the Einstein tensor, $\GN$ is
Newton's constant, $T_{\alpha\beta}$ is the energy--momentum tensor,
$\rho$ is the energy density and $u^\a$
is the velocity of observers comoving with the dust.
In the real universe, the matter cannot locally be treated
as dust everywhere, but the deviations are unlikely to be
relevant for quantities integrated over large scales, which
is what enters into the observations. For treatment of non-dust
matter, see \cite{Buchert:2001, Rasanen:2009b}.

The evolution and constraint equations can be written elegantly
in terms of the gradient of $u_\a$ and the electric and magnetic
components of the Weyl tensor \cite{Ellis:1971, Tsagas:2007},
\bea \label{gradu}
  \nabla_\b u_\a
  &=& \frac{1}{3} h_{\a\b} \theta + \sigma_{\a\b} + \omega_{\a\b} \ ,
\eea

\noindent where $h_{\a\b}$ projects orthogonally to $u^\a$.
The trace $\theta\equiv\nabla_\a u^\a$ is the volume expansion rate,
the traceless symmetric part $\sigma_{\a\b}$ is the shear tensor
and the antisymmetric part $\omega_{\a\b}$ is the vorticity tensor.
For an infinitesimal fluid element, $\theta$ indicates how
its volume changes in time, keeping the shape and the orientation
fixed, while shear changes the shape and vorticity changes the
orientation. In the FRW case, the volume expansion rate is just
$3 H$, where $H$ is the Hubble parameter.

The equations can be be decomposed into scalar, vector and tensor
parts with respect to the spatial directions orthogonal to $u^\a$.
We need only the scalar parts (we omit a scalar equation related
to the vorticity),
\bea
  \label{Rayloc} \dot{\theta} + \frac{1}{3} \theta^2 &=& - 4 \pi \GN \rho - 2 \sigma^2 + 2 \omega^2 \\
  \label{Hamloc} \frac{1}{3} \theta^2 &=& 8 \pi \GN \rho - \frac{1}{2} \sR + \sigma^2 - \omega^2 \\
  \label{consloc} \rhodot + \theta\rho &=& 0 \ ,
\eea

\noindent where a dot stands for derivative with respect to
proper time $t$ measured by observers comoving with the dust, 
$\sigma^2\equiv\ha\sigma^{\alpha\beta}\sigma_{\alpha\beta}\geq0$
and $\omega^2\equiv\ha\omega^{\alpha\beta}\omega_{\alpha\beta}\geq0$
are the shear scalar and the vorticity scalar, 
respectively. In the irrotational case, $\sR$ is the
spatial curvature of the hypersurface which is orthogonal
to $u^\a$; see \cite{Ellis:1990} for the definition in the
case of non-vanishing vorticity.

Equation \re{consloc} simply shows that the energy density is
proportional to the inverse of the volume, in other words
that mass is conserved. The second equation \re{Hamloc}
is the local equivalent of the Friedmann equation, and it relates
the expansion rate to the energy density, spatial curvature,
shear and vorticity.
The equation \re{Rayloc} gives the local acceleration.
Let us assume that the fluid is irrotational, i.e. that the vorticity
is zero. (See \cite{Rasanen:2009b} for the case with vorticity.)
As vorticity contributes positively to the acceleration,
putting it to zero gives a lower bound.
In this case, the local acceleration is always negative,
or at most zero. This is just an expression of the fact that gravity
is attractive for matter which satisfies the strong energy condition.

Cosmological distance observations imply that the expansion rate
has accelerated if we assume that the FRW relation between distance and
the expansion rate holds. Deviations from homogeneity and isotropy change
this relationship, so this conclusion does not necessarily hold in
the real universe; we discuss this in \sec{sec:light}.
(Based on direct measurements of the expansion rate, we can only say
that there has been less deceleration, not that the expansion has
accelerated.)
We can distinguish between apparent and actual acceleration.
Apparent acceleration means that when cosmological observations
are interpreted assuming that the universe is well described by
a FRW model, the expansion rate given by the FRW scale factor
has accelerated.
Actual acceleration means that the real volume 
expansion rate has really increased in time.
It is easy to understand how a different relationship between
the expansion rate and distance might lead to apparent
acceleration, but it is possible for inhomogeneities to
lead to actual acceleration as well, if we consider the average
expansion rate relevant for cosmological observations.

\subsection{The average expansion rate} \label{sec:av}

When discussing averages, the first question
concerns the choice of the hypersurface on which the average is taken.
We choose the hypersurface orthogonal to $u^\a$, which is also
the hypersurface of constant proper time $t$ measured by the observers.
(Discussion of this choice is postponed to \sec{sec:light}.)
The spatial average of a scalar quantity $f$ is its integral over
the hypersurface, with the correct volume element, divided by the volume:
\bea \label{av}
  \av{f}(t) \equiv \frac{ \int d^3 x \sqrt{^{(3)}g(t,\bar{x})} \, f(t,\bar{x}) }{ \int d^3 x \sqrt{^{(3)}g(t,\bar{x})} } \ ,
\eea

\noindent where $^{(3)}g$ is the determinant of the metric on the
hypersurface of constant proper time $t$.

Averaging \re{Rayloc}--\re{consloc}, we obtain the
Buchert equations \cite{Buchert:1999}
\bea
  \label{Ray} 3 \frac{\addot}{a} &=& - 4 \pi \GN \av{\rho} + \sQ \\
  \label{Ham} 3 \HH &=& 8 \pi \GN \av{\rho} - \frac{1}{2}\av{\sR} - \frac{1}{2}\sQ \\
  \label{cons} && \pat_t \av{\rho} + 3 \frac{\adot}{a} \av{\rho} = 0 \ ,
\eea

\noindent where the backreaction variable $\sQ$ contains the effect
of inhomogeneity and anisotropy,
\bea \label{Q}
  \sQ \equiv \frac{2}{3}\left( \av{\theta^2} - \av{\theta}^2 \right) - 2 \av{\sigma^2} \ ,
\eea

\noindent and the scale factor $a(t)$ is defined so that the
volume of the spatial hypersurface is proportional to $a(t)^3$,
\bea \label{a}
  a(t) \equiv \left( \frac{ \int d^3 x \sqrt{ ^{(3)}g(t,\bar{x})} }{ \int d^3 x \sqrt{ ^{(3)}g(t_0,\bar{x})} } \right)^{\frac{1}{3}} \ ,
\eea

\noindent where $a$ has been normalised to unity at time $t_0$,
which we take to be today. As $\theta$ gives the expansion rate
of the volume, this definition of $a$ is equivalent to
$3\adot/a\equiv\av{\theta}$. We also use the notation $H\equiv\adot/a$.

The Buchert equations \re{Ray}--\re{cons} have a slightly different
physical interpretation than the FRW equations due to the different
meaning of the scale factor.
In FRW models, the scale factor is a component of the metric,
and indicates how the space evolves locally.
In the present context, $a(t)$ does not describe local
behaviour, and it is not part of the metric.
It simply gives the total volume of a region.

Mathematically, the Buchert equations differ
from the FRW equations by the presence of the
backreaction variable $\sQ$ and the related
feature that the average spatial curvature can have
non-trivial evolution.
In the FRW case, $\sQ=0$ and $\av{\sR}\propto a^{-2}$.
This evolution of the spatial curvature follows from
the integrability condition between \re{Ray} and \re{Ham},
and it is also an independently known feature of
any homogeneous and isotropic model \cite{Rasanen:2007}.
Note that $\sQ$ can vanish even when the universe
is locally far from FRW. In other words, the FRW equations
may give a correct description of the average evolution
even if they are completely wrong for the local dynamics.
(By derivation, the FRW equations are meant to
describe local evolution.)

In general, $\sQ$ is non-zero, and it expresses
the non-commutativity of time evolution and averaging.
The backreaction variable $\sQ$ has two parts: the second term
in \re{Q} is the average of the shear scalar, which is also
present in the local equations \re{Rayloc}--\re{consloc}.
It is always negative (unless the spacetime is FRW,
in which case it is zero), and acts to decelerate the expansion.
In contrast, the first term in \re{Q}, the variance of
the expansion rate, has no local counterpart.
It may be called emergent in the sense that
it is purely a property of the average system.
The variance is always positive (unless the expansion
is homogeneous, in which case it is zero).
If the variance is sufficiently large compared to the
shear and the energy density, the average expansion rate
accelerates according to \re{Ray}, even though \re{Rayloc}
shows that the local expansion rate decelerates everywhere.

\section{Modelling backreaction}

\subsection{A two-region toy model} \label{sec:toy}

It may seem paradoxical that the average expansion rate
accelerates even though the local expansion rate
decelerates everywhere at all times. 
So let us first consider a simple toy model to understand
the physical meaning before moving on to a
semi-realistic model of the universe.
We give the punchline right away.
In an inhomogeneous space, different regions expand at different
rates. Regions with faster expansion rate increase
their volume more rapidly, by definition.
Therefore the fraction of volume in faster expanding regions
rises, so the average expansion rate can rise.
Whether the average expansion rate actually does rise depends on
how rapidly the fraction of fast regions grows relative to the
rate at which their expansion rate decelerates.

In the early universe, the distribution of perturbations of the
density, and thus of the expansion rate, is very smooth, with only
small local variations. In a simplified picture, overdense regions
slow down more as their density contrast grows, and eventually
they turn around and collapse to form stable structures.
Underdense regions become ever emptier, and their
deceleration decreases.
Regions thus become more differentiated and the variance
of the expansion rate grows.

We can illustrate this with a simple toy model where there
are two spherically symmetric regions, one underdense
and one overdense \cite{Rasanen:2006a, Rasanen:2006b}.
We consider the regions to be Newtonian, so their evolution
is given by the spherical collapse model and the underdense
equivalent, i.e. they expand like dust FRW universes
with negative and positive spatial curvature, respectively.
We denote the scale factors of the underdense and the overdense
region by $a_1$ and $a_2$, respectively.
We take the underdense region, which models a cosmological void,
to be completely empty, so it expands like $a_1\propto t$.
The evolution of the overdense region, which models
the formation of a structure such as a cluster, is given by
$a_2\propto 1-\cos\phi$, $t\propto \phi-\sin\phi$, where
the parameter $\phi$ is called the development angle.
The value $\phi=0$ corresponds to the big bang singularity,
from which the overdense region expands until $\phi=\pi$,
when it turns around and starts collapsing. The region
shrinks to zero size at $\phi=2\pi$.
In studies of structure formation, the collapse
is usually taken to stabilise at $\phi=3\pi/2$
due to vorticity and velocity dispersion, and we also
follow the evolution only up to that point. The total
volume is $a^3=a_1^3+a_2^3$. The average expansion rate
and acceleration are
\bea
  \label{Hex} \!\!\!\!\!\!\!\!\!\!\!\!\!\!\!\!\!\!\!
H &=& \frac{ a_1^3 }{ a_1^3 + a_2^3 } H_1 + \frac{ a_2^3 }{ a_1^3 + a_2^3 } H_2 \equiv v_1 H_1 + v_2 H_2 \\
  \label{accex} \!\!\!\!\!\!\!\!\!\!\!\!\!\!\!\!\!\!\!
\frac{\addot}{a} &=& v_1 \frac{\addot_1}{a_1} + v_2 \frac{\addot_2}{a_2} + 2 v_1 v_2 (H_1-H_2)^2 \ .
\eea

The average expansion rate is the volume-weighted average of
the expansion rates $H_1$ and $H_2$, as one would expect.
It is therefore bounded from above by the fastest local expansion rate.
However, from the fact that both $H_1$ and $H_2$ decrease it does not
follow that their weighted average would decrease, or that the average
expansion rate would decelerate.
This is illustrated by the acceleration equation \re{accex}.
The first two terms are the volume-weighted average,
and because the regions decelerate (or at most have zero acceleration,
in the completely empty case), it is negative.
However, there is an additional term related to the difference between
the two expansion rates, which is always positive (as long as the
regions have non-zero volume and different expansion rates).
This term arises because a time derivative of \re{Hex}
operates not only on $H_1$ and $H_2$, but also on $v_1$ and $v_2$.
In terms of the general acceleration equation \re{Ray}, the first
two terms in \re{accex} come from the average density, and the
last term is (one third of) the backreaction variable $\sQ$.

\begin{figure}
\hfill
\begin{minipage}[t!]{5.1cm} 
\scalebox{1.2}{\includegraphics[angle=0, clip=false, trim=0cm 2cm 0cm 0cm, width=0.9\textwidth]{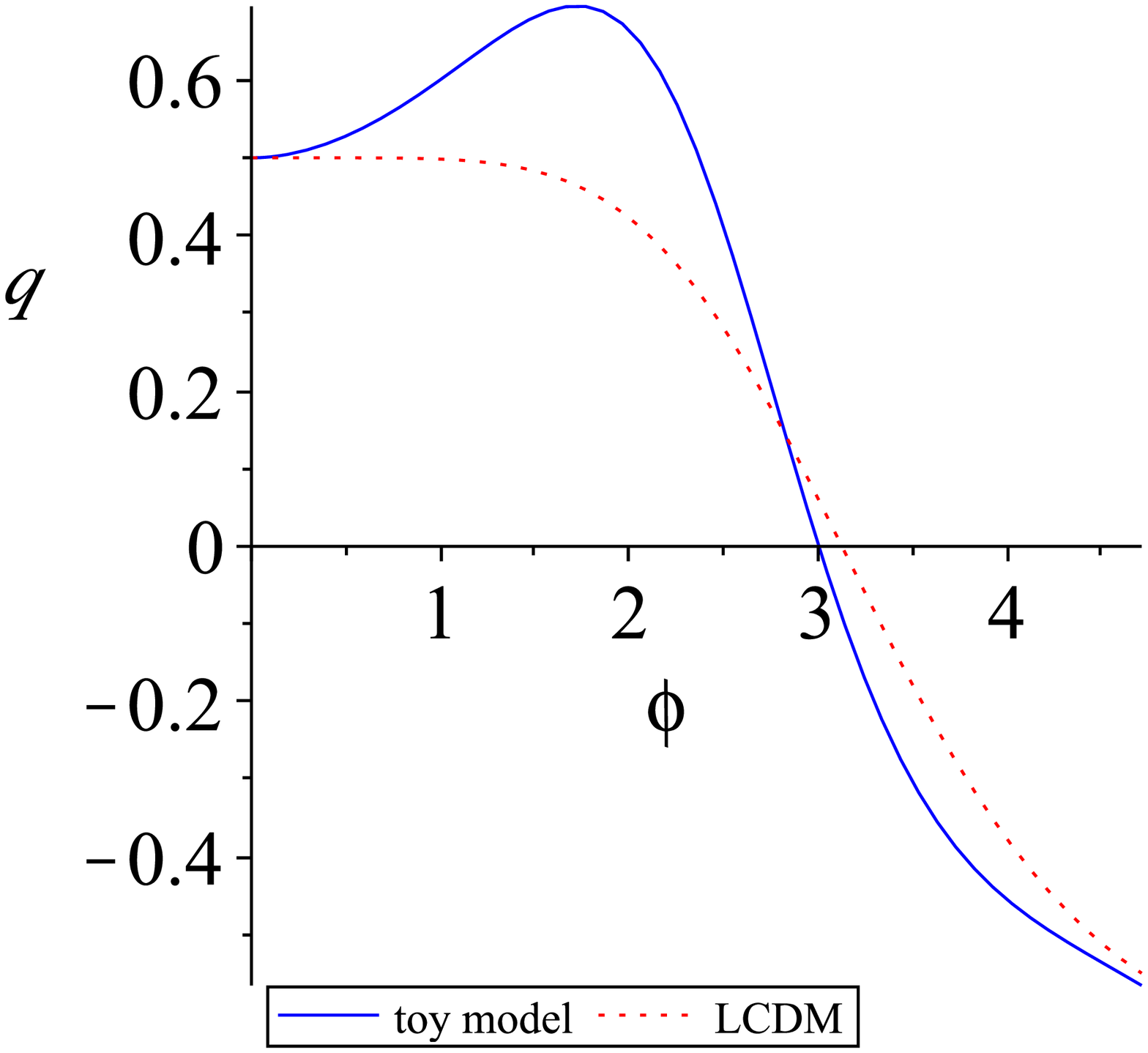}}
\begin{center} {\bf (a)} \end{center}
\end{minipage}
\hfill
\begin{minipage}[t!]{5.1cm}
\scalebox{1.2}{\includegraphics[angle=0, clip=false, trim=0cm 2cm 0cm 0cm, width=0.9\textwidth]{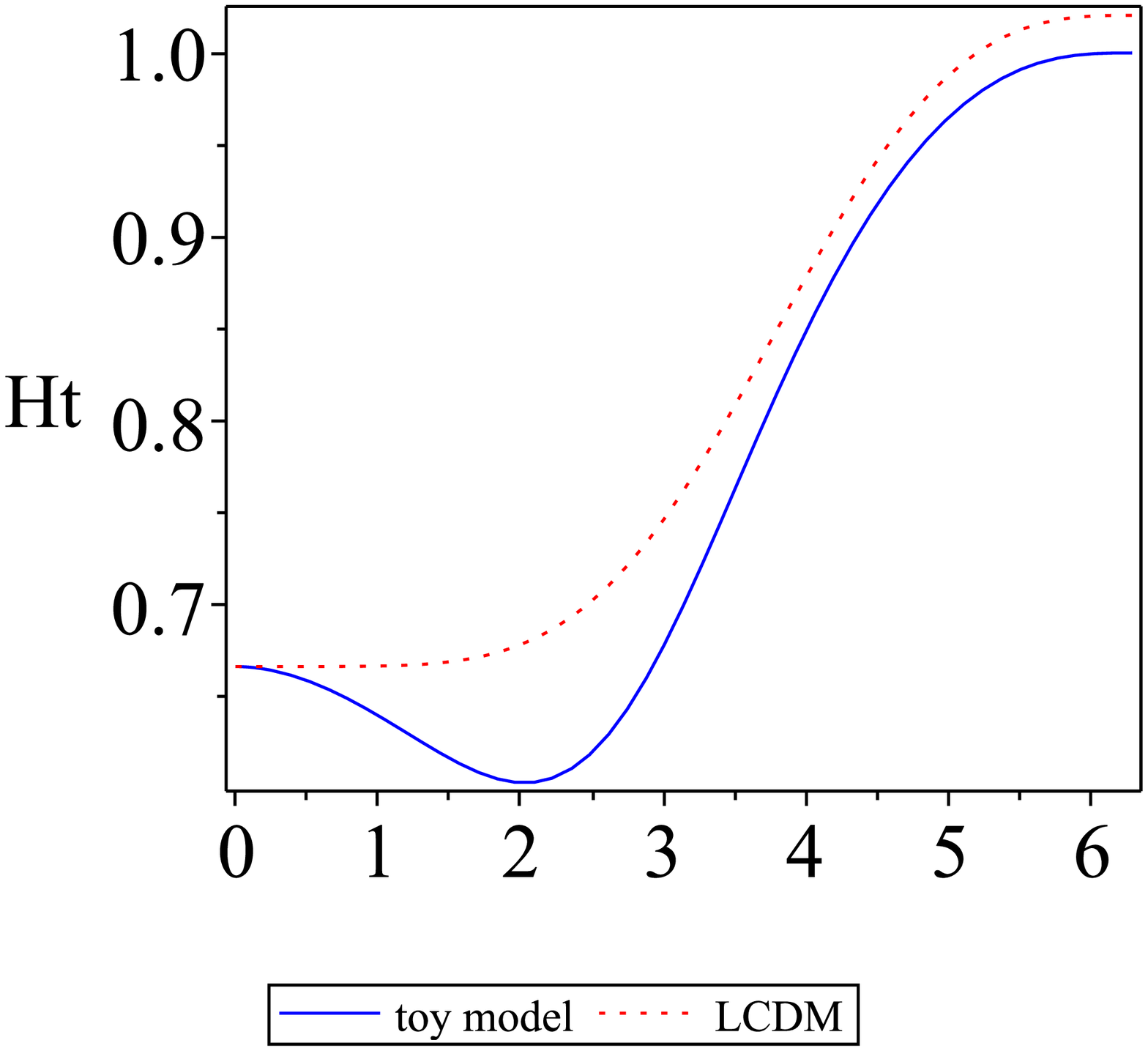}}
\begin{center} {\bf (b)} \end{center}
\end{minipage}
\hfill
\caption{The evolution of the toy model as a function of the
development angle $\phi$.
(a): The deceleration parameter $q$ in the toy model (blue, solid)
and in the \LCDM model (red, dash-dot).
(b): The Hubble parameter multiplied by time, $Ht$, in the toy model
(blue, solid) and in the \LCDM model (red, dash-dot).}
\label{fig:toy}
\end{figure}

The toy model has one free parameter, the relative size of the
two regions at some time. For illustration purposes, we
fix this by setting the deceleration parameter
$q\equiv-\addot/a/H^2$ at $\phi=3\pi/2$ to the value
of the spatially flat \LCDM FRW model with $\Omega_\Lambda=0.7$.
In \fig{fig:toy} (a) we plot $q$ as a function
of the development angle $\phi$.
We also show the \LCDM model for comparison.

The \LCDM model starts matter-dominated, with $q=1/2$.
As vacuum energy becomes important, the model
decelerates less and then crosses over to acceleration.
Asymptotically, $q$ approaches $-1$ from above as the
Hubble parameter approaches a constant value.
The backreaction model also starts with the FRW
matter-dominated behaviour, then the expansion slows
down more, before $q$ turns around and the expansion
decelerates less and eventually accelerates:
in fact the acceleration is stronger than in the \LCDM model.

The acceleration is not due to regions speeding up
locally, but due to the slower region becoming less
represented in the average.
First the overdense region brings down the expansion rate,
but its fraction of the volume falls because of the slower
expansion, so eventually the underdense region takes over
and the average expansion rate rises. This is particularly
easy to understand after the overdense region has started
collapsing at $\phi=\pi$. Then the contribution $v_2 H_2$ 
of the overdense region to \re{Hex} is negative, and
its magnitude shrinks rapidly as $v_2$ decreases,
so it is transparent that the expansion rate increases.
Note that while there is an upper bound on the expansion
rate, there is no lower bound on the collapse rate.
Therefore, the acceleration can be arbitrarily rapid,
and $q$ can even reach minus infinity in a finite time.
(This simply means that the negative expansion rate of the
collapsing region cancels becomes equal to the positive
expansion rate of the exanding region, so
$H$ vanishes in the denominator of $q$.)
This is in contrast to FRW models, where $q\geq-1$
unless the null energy condition (or the modified gravity
equivalent) is violated.
After the overdense region stops being important, the expansion
rate will be given by the underdense region alone, and the
expansion will again decelerate. Acceleration is a transient
phenomenon associated with the volume becoming dominated
by the underdense region.

Figure \ref{fig:toy} (b) shows the Hubble parameter
multiplied by time as a function of the development angle
$\phi$. This contains the same information as \fig{fig:toy} (a),
but plotted in terms of the first derivative of the scale
factor instead of the second derivative.
In the \LCDM model, $Ht$ starts from $2/3$ in the
matter-dominated era and increases monotonically without
bound as $H$ approaches a constant. In the toy model,
$Ht$ falls as the overdense region slows down, then rises
as the underdense region takes over, approaching unity
from below. The Hubble parameter in the toy model is
smaller than in the \LCDM model at all times. Because
$H$ is bounded from above by the fastest local expansion rate,
$Ht$ cannot exceed unity. This bound also holds in realistic models:
as long as matter can be treated as dust and vorticity
can be neglected, we have $Ht\leq1$ at all times \cite{Rasanen:2005},
in contrast to FRW models with exotic matter or modified gravity.
This is a prediction of backreaction.
(For discussion of vorticity and non-dust terms
in the energy-momentum tensor, see \cite{Rasanen:2009b, Buchert:2001}.)

Whether the expansion accelerates depends on how rapidly the
faster expanding regions catch up with the slower ones, roughly
speaking by how steeply the $Ht$ curve rises.
This is why the variance contributes positively to acceleration:
the larger the variance, the bigger the difference between
fast and slow regions, and the more rapidly the fast regions
take over.

\subsection{A statistical semi-realistic model} \label{sec:real}

The toy model shows how acceleration due to inhomogeneities
can occur and makes transparent what this means physically.
Acceleration has also been demonstrated with the exact spherically
symmetric dust solution, the Lema\^{\i}tre-Tolman-Bondi model
\cite{Chuang:2005, Paranjape:2006, Kai:2006}.
So there is no ambiguity: accelerated average expansion due to
inhomogeneities is possible.
The question is whether the distribution of structures in the
universe is such that this mechanism is realised.
The statement that faster expanding regions increase
their volume more rapidly makes it sound as if there would
necessarily be less deceleration (if not acceleration)
than in the FRW case. For a set of isolated regions, this is
true: eventually, the volume will be dominated by the fastest region.
However, the characteristic feature of structures in the real
universe is their hierarchical buildup. Smaller
structures become incorporated into larger ones,
and rapidly expanding voids can be extinguished by
collapsing clouds.

The non-linear evolution of structures is too complex to
follow exactly. However, because the
universe is statistically homogeneous and isotropic,
statistical properties are enough to evaluate
the average expansion rate.
The average expansion rate is determined if we
know which fraction of the universe is in which
state of expansion or collapse.
Instead of trying to find a solution for the metric
and calculating the quantities of interest from it,
we can consider an ensemble of regions from which
we can determine the average expansion rate without having to
consider the global metric.
We now discuss a semi-realistic model which
does this by extending the two fixed regions of the
toy model to a continuous distribution of regions which evolves
in time \cite{Rasanen:2008a, peakrevs}.

The starting point is the spatially flat matter-dominated FRW
model with a linear Gaussian field of density fluctuations.
Structure formation, even though complicated, is a deterministic
process. Therefore any statistical quantity at late times
is determined by the initial distribution processed by
gravity. For a Gaussian distribution, the power spectrum contains
all statistical information.
So even in the completely non-linear regime, the average
expansion rate follows from the power spectrum. The problem
is formulating a tractable model for propagating the structures given
by the initial power spectrum into the non-linear regime with gravity.
One approach, proposed in \cite{Bardeen:1986}, is to identify
structures at late times with spherical peaks in the original
linear density field, smoothed on an appropriate scale.
The number density of peaks as a function of the smoothing
scale and peak height can be determined analytically in terms
of the power spectrum.
In the original application, the correspondence between peaks
and structures was assumed to hold only for very non-linear
overdense structures: all peaks exceeding a certain
density threshold were identified with stabilised structures.
Here the idea is a bit different: spherical peaks of any
density are identified with structures having the same linear
density contrast. Troughs are identified with spherical
voids in the same way.
(As the distribution is Gaussian, the statistics of peaks
and troughs are identical.)
We keep the smoothing threshold fixed such
that $\sigma(t,R)=1$, where $\sigma$ is the root mean square
linear density contrast, $t$ is time and $R$ is the smoothing scale.
Non-linear structures form at $\sigma\approx1$, so
$R$ corresponds to the size of the typical largest structures,
and grows in time. The smoothing is just a simplified
treatment of the complex stabilisation and evolution of
structures in the process of hierarchical structure formation.

Since the peaks are spherical and isolated, and they are individually
assumed to be in the Newtonian regime, their expansion rate
is the same as that of a dust FRW universe with the same density,
as in the toy model. The volume which is neither in
peaks nor in troughs is taken to expand like the spatially
flat matter-dominated FRW model.

\begin{figure}[t]
\centering
\scalebox{0.5}
{\includegraphics[angle=0, clip=true, trim=0cm 0cm 0cm 0cm, width=\textwidth]{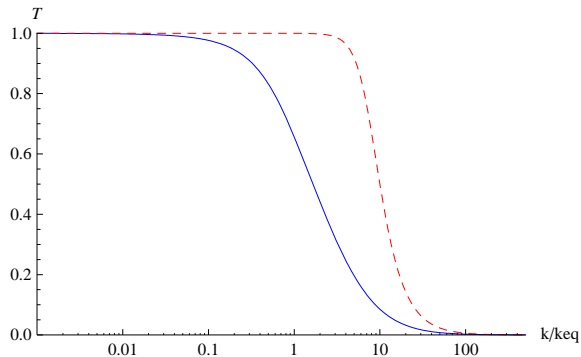}}
\caption{The BBKS (blue, solid) and BDG (red, dashed) transfer functions as a function of $k/\keq$.}
\label{fig:transfer}
\end{figure}

The peak number density as a function of time is
determined by the power spectrum, which consists of two
parts: the primordial power spectrum,
determined in the early universe by inflation or some
other process, and the transfer function, which describes the
evolution between the primordial era and the time when the
modes enter the non-linear regime. The transfer function
$T(k)$ simply multiplies the amplitude of the primordial modes.
We take a scale-invariant primordial spectrum
with the observed amplitude; small variations from
scale-invariance have little effect.
For the transfer function, we assume that dark
matter is cold, and we consider two different
approximations in order to show the uncertainty
in the calculation. The BBKS transfer function \cite{Bardeen:1986}
is a fit to numerical calculations (we take a baryon fraction
of $0.2$), and the BDG form introduced in \cite{Bonvin:2006}
is a simple analytically tractable function
with the correct qualitative features.
The transfer functions are shown in \fig{fig:transfer}
as a function of $k/\keq$, the wavenumber divided by the
matter-radiation equality scale.
Modes with $k>\keq$ enter the horizon during radiation domination,
so their amplitude is damped. The sooner they enter, the more
they are damped before the universe becomes matter-dominated,
so there is a damping tail, which falls approximately like $k^{-2}$.
Modes with $k<\keq$ enter during the matter-dominated era
and retain their original amplitude. For modes with $k\sim\keq$,
the transfer function interpolates between these two regimes.
In the BBKS transfer function, the transition is centered around
$\keq$ and is rather gradual, while in the BDG case the transition
happens a bit earlier and is more rapid. Even the more
realistic BBKS transfer function has an error of 20--30\%
compared to Boltzmann codes.

We have
\bea \label{H}
  H(t) = \int_{-\infty}^{\infty} \rmd\delta\, v_\delta(t) H_\delta(t) \ , 
\eea

\noindent where $v_\delta \rmd\delta$ is the fraction of
volume in regions with linear density contrast $\delta$
and expansion rate $H_\delta(t)$.
The correspondence between $\delta$ and $H_\delta$ is given
by the spherical evolution model (i.e. FRW evolution),
and the distribution of regions $v_\delta(t)$ is given by the
peak statistics, which is determined by the power spectrum
of the Gaussian density field. 
With the transfer function fixed, there are no free
parameters: the expansion history $H(t)$ given by \re{H}
is completely determined.
Since the primordial spectrum is scale-invariant and the smoothing
and peak identification process does not introduce a scale,
features in the expansion rate as a function of time can only come
from the turnover at the matter-radiation equality scale in the transfer
function.

\begin{figure}
\hfill
\begin{minipage}[h]{6cm} 
\scalebox{1.0}{\includegraphics[angle=0, clip=true, trim=0cm 0cm 0cm 0cm, width=\textwidth]{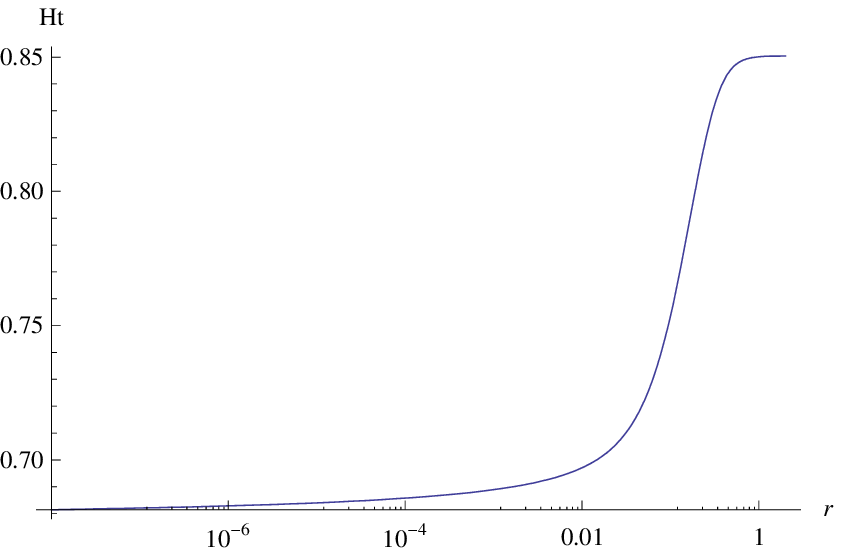}}
\begin{center} {\bf (a)} \end{center}
\end{minipage}
\hfill
\begin{minipage}[h]{6cm}
\scalebox{1.0}{\includegraphics[angle=0, clip=true, trim=0cm 0cm 0cm 0cm, width=\textwidth]{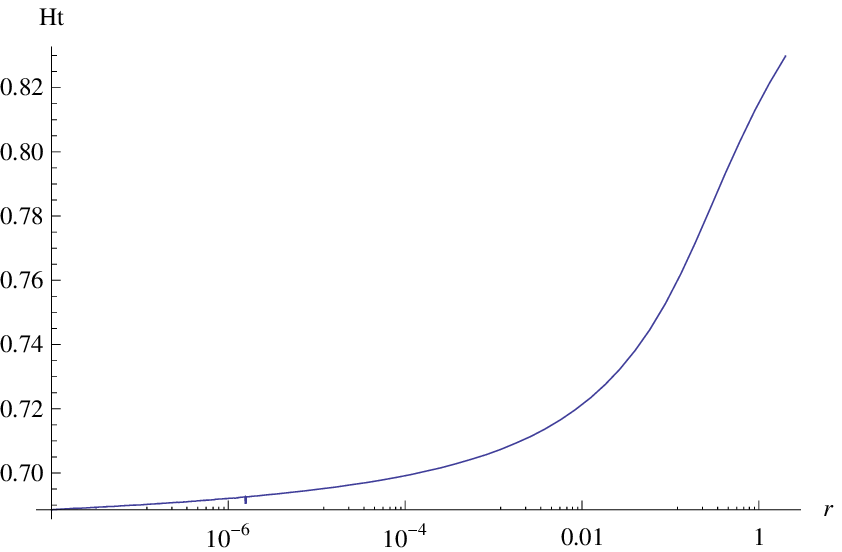}}
\begin{center} {\bf (b)} \end{center}
\end{minipage}
\hfill
\caption{The expansion rate $Ht$ as a function of $r=\keq R$ for
(a) the BDG transfer function and (b) the BBKS transfer function.}
\label{fig:Htr}
\end{figure}

In \fig{fig:Htr} we show $Ht$ as a function of
$r\equiv\keq R$, the smoothing scale relative to the
matter-radiation equality scale. Essentially, the coordinate $r$
is time as measured by the size of the largest generation of structures.
We have $Ht\approx 2/3$ at early times, as the fraction of
volume in non-linear structures is small. As time goes
on, deeper non-linear structures form, and they
take up a larger fraction of the volume.
The expansion rate grows (relative to the FRW value) slowly,
until there is rapid rise and saturation, roughly at the scale
of matter-radiation equality.
It is clear that after $r=1$, when the perturbations which
correspond to the matter-radiation equality scale collapse,
$Ht$ must settle to a constant, since the transfer function
is nearly unity, and there is no scale in the system anymore.

The matter-radiation equality scale is
$\keq^{-1}\approx 13.7\om^{-1}$ Mpc $\approx$ 100 Mpc,
using the value $\om=0.14$ \cite{Vonlanthen:2010}.
Observationally, $\sigma(t,R)\approx1$
today on scales somewhat smaller than 8 $h^{-1}$Mpc, so
$R_0\approx$ 10 Mpc. Therefore the present day happens to be
located around $r=0.1$ in the plots -- right in the transition region.
Note that nothing related to present day has been used as input in
the calculation.

\begin{figure}
\hfill
\begin{minipage}[t]{6cm} 
\scalebox{1.0}{\includegraphics[angle=0, clip=true, trim=0cm 0cm 0cm 0cm, width=\textwidth]{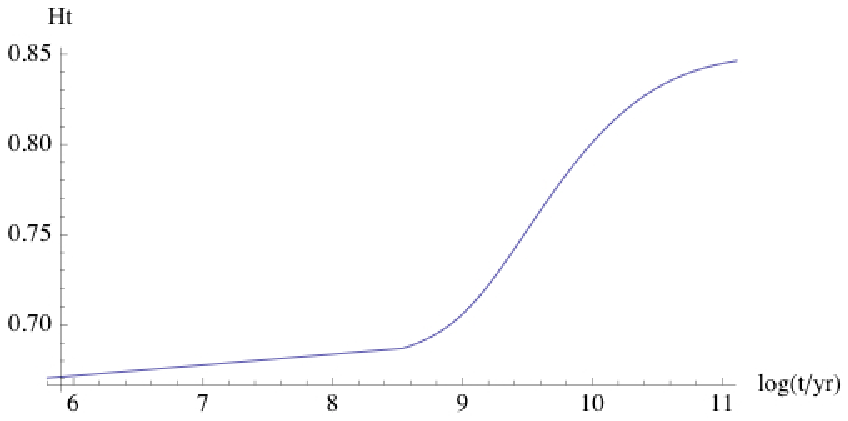}}
\begin{center} {\bf (a)} \end{center}
\end{minipage}
\hfill
\begin{minipage}[t]{6cm}
\scalebox{1.0}{\includegraphics[angle=0, clip=true, trim=0cm 0cm 0cm 0cm, width=\textwidth]{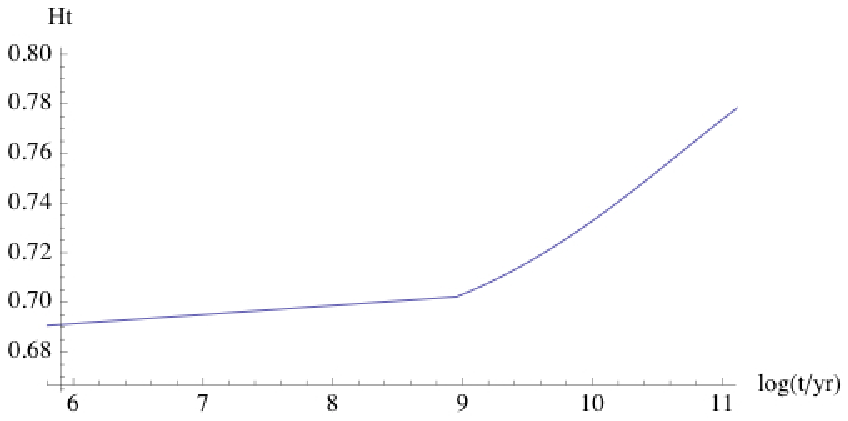}}
\begin{center} {\bf (b)} \end{center}
\end{minipage}
\hfill
\caption{The expansion rate $Ht$ as a function of time for
(a) the BDG transfer function and (b) the BBKS transfer function.}
\label{fig:Ht}
\end{figure}

It is instructive to view $Ht$ also as function of time as measured
in years. In \fig{fig:Ht}, the horizontal axis is $\log_{10}(t/$yr).
For the BDG transfer function, $Ht$ has the FRW value at
one million years, and it grows very slowly until it rises
at about a billion years, and then saturates to a value somewhat
larger than $0.8$ at some tens of billions of years. For the
more realistic BBKS transfer function, the behaviour is
qualitatively the same, but the transition is slower
and the final value of $Ht$ is smaller.
The slope of the $Ht$ curve is less steep as a function
of time than as a function of $r$, because the size of
structures grows more slowly at late times.
When plotting $Ht$ as function of the smoothing
scale, the comparison scale is $\keq$, whereas here it is the time
of matter-radiation equality, $\teq$. Now the amplitude of the
primordial perturbations also enters.
The timescale follows from the shape of the transfer function.
Perturbations which entered the horizon at matter radiation equality
reach non-linearity at $t\approx A^{-3/2}\teq\approx100$ Gyr,
where $A=3\times10^{-5}$
is the primordial amplitude and the matter-radiation equality
time is $\teq\approx 1000 \om^{-2}$ years $\approx$ 50 000 years
for $\om=0.14$. This is when the expansion rate saturates, and
it enters the transition region somewhat earlier.

As noted in \sec{sec:toy}, whether or not the expansion
accelerates is a quantitative question
related to the slope of the $Ht$ curve.
In the present case, while the expansion rate increases relative
to the FRW value, the change is not sufficiently rapid for the
expansion to accelerate, there is just less deceleration.
This is related to the fact that, unlike in the toy model,
the overdense regions play almost no role, and
the evolution of $Ht$ can be understood in
terms of the underdense voids.
At early times, voids take up only a small part of the volume,
and $Ht$ rises smoothly as their volume fraction increases.

Nevertheless, it is encouraging that the model gives a change
of the right order of magnitude in $Ht$, 15--25\%, and that
the timescale for the change comes out right.
The model involves many approximations, such as
treating structures as spherical, using an approximate transfer
function, having an artificial split between the peaks/troughs
and the smooth space, not taking into account that the Gaussian symmetry
between the overdense and underdense regions is broken in the non-linear
regime (in the present treatment, they have equal mass in at all times) and
treating the structures as isolated even for small density contrasts
and high peak number densities.
It is clear the model cannot be trusted beyond an order of magnitude.
It is also possible that a more careful statistical treatment would
reveal cancellations that significantly change
backreaction from this approximate estimate.

In order to obtain a more drastic change in $Ht$, the expansion
rate should have extra deceleration due to overdense regions
before the voids take over, as in the toy model.
(This effect is present in the model, but it is too small
to be visible in figures \re{fig:Htr} and \re{fig:Ht}.)
If the expansion were to slow down more before the voids take over,
the variance and the change in the expansion rate would be larger.
The magnitude of the change of $Ht$ is easy to
understand: if the universe were completely dominated by totally
empty voids, we would have $Ht=1$. Since not all of the volume is
taken up by voids and they are not totally empty, $Ht$ is
somewhat smaller than unity.
As noted in the case of the toy model, $Ht<1$ is a prediction
of backreaction \cite{Rasanen:2005}, assuming that matter can be
treated as dust and rotation can be neglected \cite{Rasanen:2009b}.
The constraint $Ht<1$ also means that proposals for
implementing primordial inflation using backreaction
\cite{brinfl} are unfeasible (aside from other problems,
such as generating the spectrum of primordial perturbations).

The change in the expansion can also be viewed in terms of
the deceleration parameter $q$ (see \cite{Rasanen:2008a}
for the plots). From \re{Ray} and \re{Q} we have
$q=\frac{1}{2}\Om-2(\av{\theta^2}-\av{\theta}^2)/\av{\theta}^2+6\av{\sigma^2}/\av{\theta}^2$.
We can obtain a conservative lower bound on this parameter
by taking into account $\av{\sigma^2}>0$ and $\av{\theta^2}<(3 t^{-1})^2$,
where the latter inequality follows from the fact that the local
expansion rate cannot be higher than $3 t^{-1}$.
This gives $q>\frac{1}{2}\Om-2 [(Ht)^{-2}-1]$.
For a realistic distribution of structures, the value of
$q$ is likely to be much above this bound.
For the values $\Omega_{\mathrm{m0}}=0.3$ and $H_0 t_0=0.8\ldots1$
we have $q_0>-0.98\ldots0.15$.
There is a tension between obtaining a large value of $Ht$
and a very negative value of $q$ simultaneously.
The physical reason is that in order to have $Ht$ close to unity,
a large fraction of the volume has to be in regions which are
nearly empty. This in turn means that the variance, and hence
acceleration, is smaller.
In terms of the spatial curvature, we see from \re{Ray} and \re{Ham}
that for $\Omega_{\mathrm{m0}}=0.3$ and $q_0=-0.55$,
we have $\av{\sR}_0=-6.3 H_0^2$, or
$\Omega_{R0}\equiv-\av{\sR}_0/(6 H_0^2)=1.05>1$.
The spatial curvature is large, because much of the volume
is occupied by very underdense regions.
We should note two caveats with regard to $q_0$.
First, while we have model-independent measurements of
the Hubble parameter today, determinations
of the deceleration parameter from the data are
very dependent on the assumed parametrisation of
the expansion history \cite{trans}.
For backreaction, we would expect extra deceleration
before the acceleration, and the expansion will return
to deceleration as the voids take over.
This sort of evolution is excluded by construction in
most parametrisations of the expansion history.
The data, however, does not exclude the possibility that the
expansion could have already have gone from acceleration back to
deceleration. 
There have been arguments that the observations would
in fact slightly prefer deceleration today \cite{latedec}
and extra deceleration before acceleration \cite{earlydec},
though such trends are not statistically significant in the present data.
Another caveat is that the deceleration parameter is determined
from distance observations, assuming that the relation between
distance and expansion rate is the one given by the FRW metric.
However, clumpiness changes this relation, a topic we now turn to.

\section{Light propagation} \label{sec:light}

\subsection{The choice of hypersurface}

Historically, studies of the average expansion rate
and light propagation have been somewhat disconnected.
In perturbative light propagation calculations, the change
of the average evolution has often been neglected, while
studies of the average expansion rate have usually not
made the connection to observations of light.
However, the primary quantities are the observable redshift
and distance, and averages are useful only insofar as they
give an approximate description of what is observed
\cite{Rasanen:2008b, Rasanen:2009b}.
As we discuss below, it is the requirement that the average expansion rate
describes light propagation which fixes the hypersurface of averaging.
The physically relevant averaging hypersurface cannot be
determined on abstract mathematical grounds.
This is a crucial feature, given that the averages depend on
the choice of hypersurface.
Note that the averaging hypersurface is a physical choice,
and should not be confused with choice of coordinates nor
choice of gauge \cite{Geshnizjani, Rasanen:2004, Kolb:2004}.
The derivation of the Buchert equations \re{Ray}--\re{cons} is
entirely covariant, and the result is uniquely defined in terms
of measurable quantities. It does not depend on coordinates
(indeed, it is not necessary to specify the coordinate system).
Because the treatment is non-perturbative and does not refer to
a background, there is also no question of gauge choice (which refers
to a mapping between the real spacetime and a fictitious background).

It has been argued that the procedure of averaging only scalar
quantities is somehow incomplete, and various proposals have been
put forth for averaging tensors.
The macroscopic gravity formalism \cite{Zalaletdinov},
for example, extends general relativity so that one can
map the physical manifold onto another manifold, which
is in some sense an average of the real one. 
However, the issue at hand is the effect of deviations from spatial
homogeneity and isotropy in the fixed spacetime geometry which
describes the real universe, not quantities calculated in some
other spacetime. (See section 4.1 of \cite{Rasanen:2009b}.)
Averages and ensembles can be useful for describing cosmological
observations which probe large scales because of the statistical
homogeneity and isotropy of the universe. However, it has to be 
demonstrated that they really describe observational quantities.

Almost all cosmological observations are made
along the lightcone, measuring the redshift,
the angular diameter distance (or equivalently
the luminosity distance) and other quantities
related to bundles of light rays.
In a general spacetime, these quantities are not determined
solely by expansion, and certainly not by the average
expansion rate along spacelike slices of simultaneity.
However, in a statistically homogeneous and isotropic universe
where the distribution evolves slowly, the average expansion
rate does determine the leading behaviour of the redshift
and the distance \cite{Rasanen:2008b, Rasanen:2009b}.
Considering the real observables also fixes the choice
of averaging hypersurface. We now sketch the argument for this.

\subsection{The redshift}

In a general dust spacetime, the redshift is given by
(see \cite{Rasanen:2009b} for the non-dust case)
\bea \label{z}
  1 + z &=& \exp\left( \int_{\eta}^{\eta_0} \rmd \eta \left[ \frac{1}{3} \theta + \sigma_{\a\b} e^\a e^\b \right] \right) \ ,
\eea

\noindent where $\eta$ is defined by
$\pat/\pat\eta \equiv (u^\a + e^\a) \pat_\a$, and $e^\a$
is the spatial direction of the null geodesic. If there are no
preferred directions and the change in the distribution
is slow compared to the time it takes for a light ray to
pass through a homogeneity scale sized region, the integral
over $\sigma_{\a\b} e^\a e^\b$ is suppressed.
In the real universe, if the homogeneity scale is around 100 Mpc,
then it is indeed much smaller than the timescale for the change in
the distribution, which is given by the Hubble scale
$H_0^{-1}=3000 h^{-1}$Mpc. In the early universe, structure formation
is less advanced, so the homogeneity scale is even smaller
relative to the Hubble scale further down the null geodesic.
The direction $e^\a$ changes slowly for typical light rays
\cite{Rasanen:2009b}, whereas the dust shear is correlated with
the shape and orientation of structures and changes on the
length scale of those structures. If there are no preferred
directions, over large scales structures are oriented in all
directions equally, so $\sigma_{\a\b}$ should contribute via its
trace, which is zero. Therefore the integral over
$\sigma_{\a\b} e^\a e^\b$ should vanish, up to statistical
fluctuations and corrections from correlations between 
$\sigma_{\a\b}$ and $e^\a$ and evolution of the distribution.
We can split the local expansion rate as $\theta=\av{\theta}+\Delta\theta$,
where $\Delta\theta$ is the local deviation from the average, and
similarly argue that the integral of $\Delta\theta$ is suppressed
relative to the contribution of the average expansion rate.
This suppression of the dependence on direction also explains
how the small anisotropy of the cosmic microwave background (CMB)
is not in contradiction with order unity perturbations in the
geometry \cite{Rasanen:2009a}.

Here the choice of hypersurface is important.
For the cancellations to occur, the averaging has to be done
on the hypersurface of statistical homogeneity and isotropy.
(In addition, the evolution of the distribution from one
hypersurface to another has to be slow compared to
the homogeneity scale.)
This defines the hypersurface of averaging.
In \sec{sec:av} we took the average on the hypersurface of
constant proper time of observers comoving with the matter.
Since the evolution of structures is governed by the proper
time, one can argue that this is close to the hypersurface
of statistical homogeneity and isotropy
\cite{Rasanen:2006b, Rasanen:2008a, Rasanen:2008b}.
However, these hypersurfaces will not be exactly the same,
and in the realistic case when the observer velocity is not
irrotational, the hypersurface of constant proper time is
not orthogonal to the observer velocity.
The details are thus more complicated, but 
non-relativistic changes in the velocity field which
defines the hypersurface of averaging lead only to small changes
in the averages, as long as the distribution
is statistically homogeneous and isotropic, and the averaging
scale is at least as large as the homogeneity scale \cite{Rasanen:2009b}.

Given that $\av{\theta}=3\adot/a$, we obtain $1+z\approx a(t)^{-1}$,
the same relation between expansion and redshift as in the FRW case.
This result depends on the fact that the shear and the expansion
rate enter into the integral \re{z} along the null geodesic linearly.
In contrast, the shear and the expansion rate
enter quadratically into the equations of motion \re{Rayloc}--\re{consloc}
for the geometry, so the variations do not cancel in the average,
and instead we have the generally non-zero backreaction variable $\sQ$.

\subsection{The distance}

For the angular diameter distance, we can apply similar
qualitative arguments to obtain the result \cite{Rasanen:2008b}
\bea \label{DA}
  H \pat_{\bz} \left[ (1+\bz)^2 H \pat_{\bz} \bar{D}_A \right] &\approx& - 4\pi\GN \av{\rho} \bar{D}_A \ ,
\eea

\noindent where $\bar{D}_A$ is the dominant
part of the angular diameter distance with the corrections to the
mean dropped, and the same for the redshift, $1+\bz\equiv a(t)^{-1}$.
From the conservation of mass, \re{cons}, it follows that
$\av{\rho}\propto(1+z)^3$. The distance is therefore determined
entirely by the average expansion rate $H(z)$ and the normalisation
of the density today, i.e. $\Omega_{\mathrm{m0}}$.
For a general FRW model, $\av{\rho}$ in \re{DA}
would be replaced by $\rho+p$.
So the equation for the mean angular diameter distance in
terms of $H(z)$ in a statistically homogeneous and isotropic
dust universe (with a slowly evolving distribution) is
the same as in the FRW $\Lambda$CDM model.
If backreaction were to produce exactly the same expansion
history as the $\Lambda$CDM model, the distance-redshift
relation would therefore also be identical.
This is the case even though the spatial curvature would
be large, as the spatial curvature affects the distances
differently than in the FRW case.
Note that in a general spacetime, the luminosity distance is related
to the angular diameter distance by $D_L=(1+z)^2 D_A$
\cite{Ellis:1971}, so from the theoretical point of view
it measures the same thing.

Backreaction is not expected to produce an expansion history
identical to the $\Lambda$CDM model: if the expansion accelerates
strongly, then this is likely to be preceded by extra deceleration,
unlike in \LCDM. Therefore the distances will also be different.
However, the backreaction distance-redshift relation will be
biased towards the $\Lambda$CDM model, compared to a FRW model
with the same expansion history as in the backreaction case.
The reason is that in the FRW model, the
equation for $D_A$ is modified not only by the
change in $H(z)$, but also by the change in $\rho+p$.
This may help to explain why distance observations prefer
a value close to $-1$ for the effective equation of state.

It has been pointed out that the relation between $D_A(z)$ and $H(z)$
can be used as a general test of FRW models \cite{Clarkson:2007b}.
If we measure the distance and the expansion rate independently,
we can check whether they satisfy the FRW relation.
If this is not the case, the observations cannot be explained
in terms of any four-dimensional FRW model.
(An extra-dimensional model where the four-dimensional
subspace has the FRW metric would still remain a
possibility \cite{Ferrer}.)
This holds independent of the presence of dark energy or
modified gravity, because light propagation depends directly
on the geometry of spacetime, regardless of the equations of
motion which determine it.
Similarly, we can test the backreaction conjecture that the
change in the expansion rate at small redshift is due
to structure formation without having a prediction for how
the expansion rate changes, simply by checking whether the
measured $D_A(z)$ and $H(z)$ satisfy \re{DA}.
The relation \re{DA}, which violates the FRW consistency
condition between expansion and distance is a unique prediction
of backreaction which distinguishes it from FRW models.
However, the relation between the expansion rate and the
distance should be derived more rigorously, and the expected
magnitude of the violation is unclear.

The redshift, as well as null geodesic shear and deflection
\cite{Rasanen:2009b}, should also be studied in more detail.
In particular, it would be interesting to check quantitatively
the conjecture that light propagation in a statistically
homogeneous and isotropic space with a slowly evolving
distribution of small structures can be described in terms
of the average expansion rate, and to characterise
the small corrections
\cite{Rasanen:2008a, Rasanen:2008b, Rasanen:2009b}.
The small-scale pattern depends only on the
angular diameter distance \cite{Vonlanthen:2010},
but the effects on large angular scales remain to be determined.
Extending to analysis of weak lensing in the case when
the geometry is not nearly FRW is also needed for comparing
with present and upcoming data.
Swiss Cheese models  \cite{cheese}, in particular ones
with a random distribution of structures \cite{cheeserandom},
are particularly interesting for numerical work, since the
average expansion rate and density can be different from
the FRW case, and quantities related to light can be
explicitly calculated.

It has been argued that deviations from the approximation of
treating the matter as dust would be important for modelling
observations of light because of their effect on the way clocks
run in different regions of space \cite{Wiltshire}.
Note that the dust approximation does not concern the issue
of granularity, or what should be seen as the grains of dust,
nor any fundamental aspect of general relativity.
It is simply a question of the pressure, anisotropic
stress and energy flux being subdominant to the energy density.
It seems unlikely that for a matter content of Standard Model
particles and cold or warm dark matter these quantities would
be so important in a significant fraction of space as to have
a major impact on light propagation over large scales.
And if that were the case, it is unlikely that the effects would
be captured by simply having clocks run at different rates in
different regions \cite{Rasanen:2009b}.

\section{Discussion}

\subsection{Beyond Newton}

A model is often understood better when it is
considered in a larger context, outside its domain of validity.
In particular, some special features of FRW models
are better appreciated when they are viewed as a limit
of general spacetimes with no exact symmetries.
One example is the consistency condition between
distance and expansion rate discussed above, which
is properly viewed as prediction of the FRW model
to be observationally tested rather than a fundamental
relation. Another aspect is the Newtonian limit of general
relativity -- or more properly, the relation between
Newtonian gravity and general relativity.

Quantifying backreaction analytically or via an improved
statistical model similar to the one discussed in
\sec{sec:real} is difficult because structure formation
is by definition a non-linear process.
However, the details of the evolution of non-linear structures
starting from small perturbations in the linear
regime are routinely studied numerically in cosmological
N-body simulations.
The problem is that the simulations use Newtonian gravity
with periodic boundary conditions.
In Newtonian gravity, the variance and the shear cancel in
the backreaction variable $\sQ$ given in \re{Q}, up to total derivatives
which can be written as boundary terms \cite{Buchert:1995}.
Boundary terms of course vanish for periodic boundary conditions.
However, using a large simulation and considering boxes of the size
of the observable universe
would not help the situation. Total derivative terms represent a flux,
and due to statistical homogeneity and isotropy, the integrated flux
over the boundary should vanish (up to statistical fluctuations),
as otherwise there would be a preferred direction.

In general relativity, the backreaction variable
$\sQ$ does not reduce to a boundary term,
and the average expansion rate of a volume depends on the behaviour
everywhere in the volume, not just on the boundary.
In contrast, the Newtonian evolution is sensitive to boundary
conditions, even for infinitely far away boundaries.
This is related to the fact that the Poisson equation
is elliptic and not hyperbolic, so the Newtonian system
of equations not have a well-posed initial value problem.
This is one aspect of the qualitative difference between general
relativity and what is called Newtonian cosmology.
The small-velocity, weak field limit of general relativity is not
Newtonian gravity, as demonstrated by the existence of Newtonian
solutions which are not the limit of any general relativity solution
\cite{Ellis:1971, Senovilla:1997}.
Rather, it is a theory with new degrees of freedom and
additional constraints compared to Newtonian gravity
\cite{Ellis:1971, Ellis:1994, Ehlers:1999, Szekeres:2000, Rasanen:2010a}.
The formulation of this limit of general relativity in the
cosmological setting with non-linear perturbations is an open issue.

In Newtonian gravity, the feature that inhomogeneities do not change
the average expansion rate in a statistically homogeneous and isotropic
universe can be understood in terms of energy conservation.
In the exactly homogeneous and isotropic case, the Newtonian Friedmann
equation (multiplied by $a^2$) can be interpreted as stating
that the kinetic energy plus the potential energy is constant.
The relativistic Friedmann equation is mathematically identical,
but has a different physical interpretation, with the
constant energy replaced by the spatial curvature term.
This correspondence does not hold beyond the FRW case.
In Newtonian gravity, the total energy is conserved even
when the system is inhomogeneous and anisotropic, as long
as the system is isolated (i.e. the boundary terms in $\sQ$ vanish).
However, in general relativity, there is no conservation
law for the average spatial curvature, and
$a^2\av{\sR}$ is in general not constant.
The FRW model is rather special in that the relativistic
spatial curvature behaves exactly like the Newtonian energy.

In building a statistical model to evaluate backreaction
effects to improve on the semi-realistic treatment discussed
in \sec{sec:light}, it is important to make sure that it is
consistent with the relativistic evolution equations and
constraints, instead of the Newtonian ones.
(Similarly, for numerical studies, one should include the
relevant relativistic degrees of freedom in the simulation.)
For example, if the peak model of \sec{sec:light} were to be
considered in a Newtonian setting, we would have to take
into account that the peak identification process does
not conserve the Newtonian energy (or correspondingly the
relativistic spatial curvature).
Taking this constraint into account would completely cancel
the effect seen in the model.

While there is no such exact cancellation in general relativity
(in the non-linear regime; we discuss the linear case below),
it has been argued that for solutions relevant for the real
universe there is nevertheless a strong cancellation
between the variance of the expansion rate and the shear
in the backreaction variable $\sQ$ given in \re{Q}
\cite{Paranjape, Mattsson:2010a, Mattsson:2010b}.
However, the models which describe a single
spherically symmetric structure are not realistic.
They only show that in some models backreaction is small,
just as it has been demonstrated that it is large for other
spherical solutions \cite{Chuang:2005, Paranjape:2006, Kai:2006}.
In \cite{Mattsson:2010b}, a Swiss Cheese construction was
also considered. However, the individual holes either
have zero shear everywhere except at an infinitely thin shell
(where both the shear and the expansion rate diverge),
or consist of two regions, one of which has zero variance and
zero shear, and the other has zero variance but non-zero shear.
Both cases are unrealistic, the first case because the shear
and the expansion rate should remain bounded, and the second
because in general non-zero shear is accompanied by non-zero
variance of the expansion rate.

\subsection{Beyond linearity}

It has been argued that backreaction is small, because metric
perturbations remain much smaller than unity even when the
density perturbation becomes non-linear.
However, it is not clear whether metric perturbations
indeed remain small. Furthermore, the observables depend
not only on the metric but also on its derivatives, which
have non-linear fluctuations. See \cite{Rasanen:2010a} for discussion.
A recent paper making this argument is \cite{Baumann:2010},
but there the averages are taken over the background space,
not the physical volume, so they commute with time derivatives
and backreaction is suppressed by construction.
(This also means that the results depend on the chosen
coordinate system and gauge.)
A much more interesting analysis is \cite{Green:2010}, 
where the background and perturbations are carefully defined;
the approach deserves further study.
Also, whether the metric can be written in the perturbed FRW
form if backreaction is important is not yet clear, and should
be considered.

It has also been suggested that the effect of backreaction
could be encapsulated in a change of the evolution of the
FRW scale factor. The idea is that backreaction is simply
a matter of taking into account the effect of structures on the
choice of a FRW background.
It can be unambiguously said that this is not the case.
If backreaction is important, the universe cannot
described by the FRW metric. For example, the relation between
the distance and the expansion rate discussed in \sec{sec:light}
which follows directly from the FRW metric is in general violated
\cite{Rasanen:2009b}.

Ultimately, the relevant question is not in which form the metric
can be written, but what happens to physical quantities.
As noted earlier, in the real universe the variation
in the local expansion rate is of the same size as the
observed change in the average expansion rate, and any
realistic metric has to reproduce this fact.
The key issue is how the slow and fast expanding regions
add up and whether the variations cancel in the average.
In linear theory, and in Newtonian gravity, the cancellation
holds, but this is not true of non-linear general relativity.

\section{Summary}

The formation of non-linear structures at late times affects
the expansion of the universe and light propagation.
This may explain the observed late-time failure of the
predictions of homogeneous and isotropic models with ordinary
matter and gravity.

Clumpiness can lead to accelerated expansion
\cite{Rasanen:2006a, Rasanen:2006b, Chuang:2005, Paranjape:2006, Kai:2006}.
The observed timescale of 10 billion years and the right order
of magnitude for the change of the expansion rate emerge
from the known physics of structure formation in a
semi-realistic model \cite{Rasanen:2008a, peakrevs}
(though the model does not have acceleration, only less deceleration).
However, the model cannot be trusted beyond an order of
magnitude, and it is possible that a more detailed study
will reveal cancellations which suppress the effect.
The physical explanation is simple: at late times, the
universe becomes dominated by underdense voids, since they
expand more rapidly than their surroundings, so the average
expansion rate rises.

The basics of the change in light propagation due to structures
and how it is related to the average expansion rate
is understood \cite{Rasanen:2008b, Rasanen:2009b}.
Demanding that the average quantities give an approximate
description of light propagation also fixes the hypersurface
of averaging as the one of statistical homogeneity and isotropy.
The redshift and the average expansion rate are related in the
same way as in FRW models, provided the distribution of structures
is statistically homogeneous and isotropic and evolves slowly.
In contrast, the relation between the average expansion rate
and the angular diameter distance is different from the FRW case.
This is a unique prediction which makes it possible
to distinguish backreaction from FRW models with dark energy
or modified gravity.
Details of light propagation remain an important venue for investigation.
Interesting issues include understanding the large angle CMB anisotropy
and weak lensing in a setting where the spacetime is not assumed to be
close to FRW.

An important topic which remains to be properly addressed is the
role of non-Newtonian aspects of general relativity \cite{Rasanen:2010a}.
In Newtonian cosmology, backreaction reduces to a boundary
term, and is therefore suppressed for a statistically
homogeneous and isotropic distribution \cite{Buchert:1995}.
However, this is not the case in general relativity, and
deriving the cosmological limit of general relativity in the case
when the metric is not close to FRW is an open problem.
Making sure that the relevant non-Newtonian aspects of general
relativity are taken into account is a central issue in going
from a semi-realistic model of backreaction to a fully reliable treatment.

There has been much progress in understanding backreaction
during the last dozen years.
The backreaction conjecture that the failure of the homogeneous
and isotropic models with ordinary gravity and matter is due to
the known breakdown of homogeneity and isotropy related to
structure formation remains a plausible possibility.
Directions for further study are clear, and a lot of work
remains to be done before we know whether the conjecture
is true or false, and if it is true, how to precisely
quantify the effect.
Until this question has been answered, we do not know whether
new physics is needed to explain the observations, or if they can be
understood in terms of a complex realisation of general relativity.


\end{document}